\documentclass[12pt]{article}

\usepackage{epsfig}
\usepackage{graphicx}
\usepackage[numbers,sort&compress,merge]{natbib}
\newcommand{\sig}[1]{$e \quad \mu \,\tau \, \tau \, b\:$}

\newcommand{\bea}{\begin{eqnarray}}
\newcommand{\eea}{\end{eqnarray}}

\begin{document}
\title{ \bf \large Dilepton Signatures of the Higgs Boson with Tau-jet Tagging }
\author{Pankaj Agrawal, Somnath Bandyopadhyay and Siba Prasad Das\footnote{email:
agrawal@iopb.res.in, somnath@iopb.res.in and spdas@iopb.res.in}\\
Institute of Physics \\ Sachivalaya Marg, Bhubaneswar, Orissa, India 751 005}
\maketitle
\begin{abstract}

We consider the process $p p \to t {\bar t} H$. 
This process can give rise to many signatures of the 
Higgs boson. The signatures can have electrons, muons 
and jets. We consider the signatures that have two 
electrons/muons and jets. Tagging of a tau jet and a 
bottom jet can help reduce the backgrounds significantly.
In particular, we examine the usefulness of
the signatures ``isolated 2 electrons/muons + a bottom jet + 
a tau jet'', ``isolated 2 electrons/muons + 2 tau jets'',
``isolated 2 electrons/muons + 2 bottom jets + a tau jet'', and 
``isolated 2 electrons/muons + a bottom jet + 2 tau jets''. 
We find that signatures with two tau jets are useful. The signatures
with one tau jet are also useful, if we restrict to same-sign
electrons/muons. These requirements reduce the backgrounds
due the process with Z-bosons + jets and the production of a 
pair of top quarks. We show that these signatures may be visible 
in the run II of the Large Hadron Collider. 

\end{abstract}

\newpage

\section{Introduction}

     The Higgs mechanism of the Standard Model (SM) now seems to have
    been validated with the discovery of a Higgs boson like neutral
    scalar particle. The strong evidence has been presented by
    the both ATLAS \cite{atlashiggs}  and CMS \cite{cmshiggs} Collaborations
    on the basis of the data taken in run I (2009-12) at the  Large Hadron Collider (LHC).
    Because of the appearance of the signal in multiple channels, as seen by both 
    collaborations, there is little doubt that the Higgs boson of the
    SM has been found. All channels of the discovery 
    suggest a mass of about 125 GeV for the particle.

     The LHC is now on a long shut-down to improve the luminosity
     and the centre-of-mass energy. When it restarts to take data in 
     2015, one of its major goals would be to measure the couplings of the newly 
     discovered Higgs boson to all the SM particles. This
     is specially important because of the prediction of the 
     existence of scalar particles, sometime with properties similar
     to that of the SM Higgs boson, in various extensions
     and modifications of the standard model. To do so, one will need to
     identify the particle through multiple processes and measure the 
     couplings of the scalar particle with various 
     other SM particles. These couplings determine the branching
     ratios of the decay channels and also the production cross sections.
     Identification of the scalar particle through multiple processes
     will allow us to measure the couplings and confirm that
     the scalar particle is indeed the SM Higgs boson.

       In this letter, we consider the production of the Higgs boson in association with a  
     top-quark pair $p p \to t {\bar t} H$ \cite{cmstth,pahiggsbb}, with its subsequent
     decay into a tau-lepton pair or $ W W^{*}$. As of now the Higgs boson has been
     primarily looked through its gluon-fusion production mechanism and then
     decay into channels $H \to \gamma \gamma $\cite{atlashiggs,cmshiggs,hsmlhc}, 
     $W W^{*}$ \cite{Chatrchyan:2013iaa,Kao:hww}, $Z Z^{*}$ \cite{atlashwwsm}, ${\rm and} \; 
     \tau \tau$ \cite{Aad:2012mea,Chatrchyan:2011nx,Chatrchyan:2012vp,atlasHtautau}.
     Various production mechanisms and the decay channels of the Higgs boson
     give rise to many signatures. Some of these signatures have already been
     discussed in the literature \cite{Dittmaier:2011ti,
     Englert:2011iz,taupibonn,dutta,Baglio:2011xz,Curtin:2013zua,
     Craig:2013eta,Onyisi:2013gta,hww,Carmona:2012jk,Azatov:2013xha,
     Buckley:2013auc,Ellis:2013yxa,Englert:2014uua,Greljo:2014dka,
     Agrawal:1998hp,Agrawal:1989mr,Agrawal:1994zx}.
     In this letter, we focus on those signatures which have two electrons/muons
     (i.e., two electrons, or 2 muons, or one electron and one muon) and 
     jets in the final state. These jets can be initiated by a light quark/gluons,
     a bottom quark, or a hadronic decay of a tau lepton (tau jet). It is experimentally
     possible to tag a jet from a bottom quark or a tau lepton. Such
     tagging helps in reducing the strong interaction backgrounds. One major source
     of the backgrounds is the production of a pair of top quarks with or without 
     additional jets. One strategy to reduce this background would be
     to restrict the signature events to same-sign electrons/muons. We show
     the usefulness of this strategy, specially when only one jet is tagged as a tau
     jet.

          In the next section, we will discuss the signatures that we consider.
     In the section 3, we will discuss the backgrounds to these signatures.
     In the section 4, we will present numerical results and some discussion.
     In the last section, we will conclude.

\section{Signatures}

       We are considering a general class of signatures ``2 electrons/muons + jets''.
      As we see from Table 1, without any tagging of the jets, the backgrounds due
      to $Z$ bosons + jets and $t {\bar t}$ + jets processes would overwhelm the
      signal. Therefore, to reduce the backgrounds, we are focusing on the signatures 
      with two electrons/muons and at least two tagged jets. Since the top-quark
      background events always have bottoms jets, so to reduce it we will require
      at least one jet to be tagged as the tau jet. These signatures occur, when after
      the production of $t {\bar t} H$, the Higgs boson decays into a 
      tau-lepton pair or $WW^{*}$. 
      With these considerations, at least one
      of the top quark accompanying the Higgs boson decays semileptonically.
      The possibility of a top quark decaying into jets leads to an
      increase in the signal events, relative to when we have more than
      2 electrons/muons in a signature. For the Higgs boson  with a mass 
      of $120 - 130\,$ GeV, the tau-lepton decay mode has a branching ratios 
      of $5 - 7$ percent; the W-boson decay mode has a branching ratio
      of $14-30 \%$. When a tau lepton decays into hadrons, 
      it can manifest itself as a jet -- tau jet. This jet
      has special characteristics. It is narrow and has very few hadrons. 
      Its narrowness is due to the
      low mass of the tau lepton; it has few hadrons because a tau lepton
      decays into mostly 1 or 3 hadrons. These properties of
      a tau jet help us to distinguish it from a quark/gluon jet.
      There is usually a $25-50\%$ efficiency to tag a tau jet. The
      probability of a light quark/gluon jet to mimic a tau jet
      can be taken to be $1-0.1\%$ \cite{kaoltt,cmsepstau,Bagliesi:2007qx}.
      A bottom jet is broader than a light quark/gluon jet and
      has more particles. It can mimic a tau jet less often. A
      bottom jet can be identified with a probability of about
      $50-60\%$, while other jets can mimic it with a probability of about
      one percent \cite{cmsepsb,mdadspd,Lehmacher:2008hs}.

     To manage the background and at the same time to keep the signal
     events to a sufficiently high level, we are analyzing the signatures 
     ``2 electrons/muons + a tau jet + a bottom jet'',
    ``2 electrons/muons + two tau jets'',``2 electrons/muons + a tau jet + 
     two bottom jet'', and ``2 electrons/muons + two tau jet + a bottom jet''. In 
     the signal, the bottom jets appear due to the decay of the top quarks; a tau jet can 
     occur due to the decay of the Higgs boson, or the decay of the on-shell/off-shell
     W bosons from the Higgs boson or the top quarks. Electrons/muons can appear 
     due to a decay chain of the Higgs boson, or the decay of the top quarks.
     Presence of electrons/muons in a signature is important to reduce the 
     background. Recently, we had considered the signatures with three or four
     electrons/muons \cite{Agrawal:2013owa}. We saw that the presence of a bottom 
     jet with four electrons/muons and the presence of one additional tau jet and a bottom 
     jet with three electrons/muons help in keeping the background low 
     enough to be able to detect the signal. 

       In the case of two electrons/muons, as we will see, it will be useful
     to have either at least two tau jets or one tau jet with only
     same sign electrons/muons in the signatures.  Either of these two
     strategies will reduce the signal events, but will reduce the backgrounds
     even more. We can have same sign charged leptons in the signature because
     there are three/four on-shell/off-shell  W boson in the production and decay
     chains under considerations. Two of these W-bosons can produce the same-sign
     electrons/muons. Sources of off-shell W-bosons can be tau-lepton, which
     can come from the decays of the Higgs boson, the top quark, the W-boson,
     or the Z-boson. This strategy of observing same-sign leptons will 
     significantly reduce the large background
     from the production of a top quark pair with or without jets and Z + jets.
     Z + jets backgrounds are significantly reduced or eliminated due to the
     the tagging of at least 2 jets as tau and/or bottom jets. This tagging
     also reduces the top-quark pair production background to the same-sign
     lepton signatures. This is discussed more in the next section.

\section{Backgrounds}

       All the signatures under consideration will get contribution
       from the signal events, i.e. the production of the Higgs boson, and
       other SM processes which do not have a Higgs boson.
       To establish the viability of the
       signatures for signal detection, we shall first identify the main background processes
       and then estimate their contributions. We will consider both
       types of the backgrounds: direct backgrounds and mimic backgrounds. 
       In the case of the direct background, the background processes produce events
       similar to the signal events. They have same particles as in
       the signal. On the other hand, mimic backgrounds
       have jets, which can mimic (fake) a tau jet, a bottom jet, or
       even an electron/muon. These mimic probabilities are usually 
       quite small -- less than a percent. So even if a background
       has large cross section, it becomes smaller when folded
       with mimic probabilities. Tagging efficiencies and mimic probabilities
       were discussed in the last section. 

       One important type of background occurs when a B-meson in a bottom jet
       decays into an electron/muon and this lepton is away from the jet. This
       leads to an extra lepton in the event. Possibility of such backgrounds
       has been explored in the literature \cite{Sullivan:2010jk}. As we 
       have argued \cite{Agrawal:2013owa},
       such backgrounds which can occur due to the top quark production is
       not significant for the signatures under consideration. This is mainly
       due to two facts -- (1) we have at least one tau jet in the signatures,
       so backgrounds are to be folded with the tau jet mimic probability; this
       reduces the backgrounds significantly, (2) the electrons/muons in our
       signatures are hard and have same minimum transverse momentum as the bottom
       jet from which they might have separated; the $p_{T}^{\ell, b} > 20 \;$ GeV.
       Let us now discuss the backgrounds to the signatures.

       \begin{enumerate}

       \item ``2 electrons/muons + a tau jet + a bottom jet'': There are many 
       processes which can be backgrounds. The source of major direct
       backgrounds is the process $t {\bar t} Z$.
       The main sources of mimic backgrounds are:  $t {\bar t}, WZ \, +\;{\rm jet}, 
       t {\bar t} W, Z \, + \, 2 \;\;{\rm jets}, WW \, + 2\;\; {\rm jets}$. 
       We are not considering backgrounds when a jet mimics an electrons/muons. 
       Such mimic backgrounds are not significant because of the very small 
       probability of a light jet to mimic an electron/muon, about 
       $10^{-4} - 10^{-5}$ \cite{atlasjinst,fakelepton,jetlep, Aad:2012xsa,Aad:2010ey}.

         Among the direct backgrounds, the most significant backgrounds would be due
       to the production of $ t {\bar t} Z$ and subsequent decay
       into leptons. Because of
       similar structure,  $ t {\bar t} Z$ will always be a significant background
       to the signal. This background can be reduced by requiring
       appropriate $M_{\ell_1 \ell_2}$ to be away from the mass of the Z-boson. But the
       background when a Z-boson decays into a tau-lepton pair and the subsequent decay
       of the tau-leptons into electrons/muons cannot be reduced in this way. 
       The major mimic background is the production of a top-quark pair. Even
       with the folding of mimic probabilities, it remains large enough to
       make the signature almost not useful. However, when we consider the
       subset of events with same-sign electrons/muons, this signature
       becomes quite viable. This is because now the $t {\bar t}$ process is
       no longer a significant background.

       \item ``2 electrons/muons + two tau jets'' : In this case,
       the direct backgrounds are the processes $t {\bar t} Z, WWZ, ZZ$.
       The main sources of mimic backgrounds are:  $t {\bar t}, WZ + \;{\rm jet}, 
       t {\bar t} W, Z + \, 2 \;{\rm jets}, WW + \,2 \; {\rm jets}$. Presence of
       two tau jets will be crucial to reduce the mimic backgrounds.

       \item ``2 electrons/muons + a tau jet + 2 bottom jets'': 
       The source of major direct backgrounds is the processes $t {\bar t} Z$.
       The main sources of mimic backgrounds are:  $t {\bar t} + \;{\rm jet} , 2 WZ + \;{\rm jet}, 
       t {\bar t} W, Z + \, 3 \;{\rm jets}, WW +\,  3 \;{\rm jets}$. These backgrounds are 
       similar to that of the first signature, except that some mimic backgrounds 
       have an extra jet.

       \item ``2 electrons/muons + a bottom jet + 2 tau jets'': 
       The sources of major direct
       backgrounds are the processes $t {\bar t} Z, WWZ, ZZ$.
       The main mimic backgrounds are:  $t {\bar t} + \;{\rm jet} , 2 WZ + \;{\rm jet}, 
       t {\bar t} W, Z + 3 \;{\rm jets}, WW + 3 \;{\rm jets}$. These backgrounds are 
       similar to that of the second signature, except that some mimic backgrounds 
       have an extra jet.

       \end{enumerate}

\section{Numerical results and Discussion}

         In this section, we are presenting numerical results and discussion
         of the results. The signal and
	 the background events have been calculated using ALPGEN (v2.14) \cite{alpgen} 
         and its interface with PYTHIA (v6.325) \cite{pythia}. Using ALPGEN,
         we generate parton-level unweighted events. Using the PYTHIA interface,
	 these events are then turned into more realistic events by hadronization,
	 initial and final state radiation. We have applied  following kinematic cuts:
  
$$
p_T^{e, \mu,j} > 20 \; \rm{GeV},\: |\eta^{e, \mu,j}| < 2.5,\; R(jj,\ell j, \ell\ell) > 0.4.
$$
         We are presenting results for the three different values of 
	 $M_H$ -- 120, 125 and 130 GeV. We have used the default values 
	 for the parameters including renormalization and factorization scales. 
         For the parton distribution functions, we have used  CTEQ5L \cite{cteq5}
	 distribution. We have chosen the center-of-mass energy of 14 TeV and 
	 integrated luminosity of $100 \;$ fb$^{-1}$. The mass of the top 
	 quark is 174.3 GeV. 

          We are presenting the results for four signatures:
         ``2 electrons/muons + a tau jet + a bottom jet'',
         ``2 electrons/muons + two tau jets'',``2 electrons/muons + a tau jet + 
         two bottom jet'', and ``2 electrons/muons + two tau jet + a bottom jet''. 
         For the bottom jet,  we have used the identification
         probability of $55\%$ \cite{cmsepsb,mdadspd}. 
	 For other jets to mimic a bottom jet, we  use the probability of $1\%$.
         For a tau jet, we consider two cases. This is because of a trade-off 
	 between higher detection efficiency and higher rejection of the 
	 mimic-jets. In the first case of LTT, low tau-tagging, we have 
	 taken the low value for the tau-jet identification, $30\%$, and low 
	 mimic rate of $0.25\%$ \cite{kaoltt}. The second case of HTT 
	 \cite{cmsepstau}, high tau-tagging, 
	  has high identification rate of $50\%$ and higher mimic rate of $1\%$. 
	  To reduce the Z boson related backgrounds, we have required the missing
	  transverse momentum to be more than 25 GeV and applied a cut on the mass of 
	  same-flavour and opposite-sign lepton pair by requiring 
	  $|M_{\ell_1 \ell_2} - M_Z| > 15 \;\; {\rm GeV}$.
          We have smeared the jet/lepton energies using the energy resolution function
\bea \label{smearing}
{\Delta E \over E} = {a \over \sqrt{E}}
\oplus b,
\eea
where $\oplus$ means addition in quadrature. For the jets $a = 0.5$ and $b = 0.03$. For the
electrons/muons we take $a= 0.1$ and $b=0.007$. Since we are not using the mass of two or
more jets, inclusion of jet energy resolution does not affect the results significantly. 
Lepton energy resolution is quite good, so the results are also not significantly impacted.

	   In Table 1, we display results with only basic kinematic cuts with 
	   the observation of only two electrons/muons. The table has results for
	   the signal events and various possible backgrounds. There are two cases
	   of same-sign (SS) electrons/muons and opposite-sign (OS) electrons/muons.
	   These events may or may not have a tau or a bottom jet. This table 
	   illustrates the importance of jet tagging and observing same-sign
	   electrons/muons. First we note that there is marginal differences
	   in the two-electrons and two-muons events. This is primarily statistical,
	   i. e., due to the finite event sample. We also notice large backgrounds
	   due to Z boson processes and top-quarks only processes. A missing $p_T$
	   cut and a cut on the mass of the lepton pair will help in reducing
	   these backgrounds. Fig 1 illustrates the importance of the missing $p_T$
	   cut. We also notice the virtual elimination of the background
	   due to a top-quark pair production for the same-sign electrons/muons. 
	   However, it will come at the cost
	   of reducing the signal events by a factor of about 3. In the case of
	   only one tau jet in the signature, one will have to adopt this strategy.
	   For the two tau jets case, the extra rejection factor, due to the
	   observation of the second tau jet, can reduce the
	   backgrounds by about a two orders of magnitude, so the restriction
	   to same-sign electrons/muons is not necessary.

        In the Tables 2-5, we present results for various signatures for the integrated
        luminosity of $300 \;$ fb$^{-1}$. This is the expected luminosity for the run II.  
        We have included only the major backgrounds.  
        We have also taken into account Next-to-leading-order (NLO) contributions to the
        signal and background processes. To do so, we have multiplied the leading-order (LO)
        results by appropriate K-factors. The K-factor is taken as 1.20 for
        the $t {\bar t}H$ \cite{Beenakker:2002nc} process; the K-factors 
         for the $t {\bar t}Z$ \cite{Lazopoulos:2008de},
	 $t {\bar t}W$ \cite{Campbell:2012dh}, and $ZZ$ \cite{Campbell:1999ah} are taken 
         to be 1.35. The K-factor for the  $WZ + {\rm jet}$ \cite{Campanario:2010hp}
	 is chosen as 1.3, while for the $WWZ$ \cite{Hankele:2007sb} production, it is 1.7.
         For the processes $t {\bar t}$  \cite{ttnlo} and $t {\bar t} + \; {\rm jet}$ \cite{Dittmaier07}, 
         K-factors are taken to be
         1.5 and 1.4 respectively. For the $Z + 2 \; {\rm jet}$, the K-factor is
         1.3 \cite{Campbell:2002tg}. Because of the 
	 smaller K-factor for the signal, as compared to the backgrounds, its inclusion
         increases the significance only marginally.

\vskip 0.3in
\begin{center}
%%{\tiny{
{\scriptsize{
\begin{tabular}{||c||c|c|c|c|c|c||c|c|c||} \hline
& \multicolumn{2}{c|}{$e \mu$}& \multicolumn{2}{c|}{$ee$}& \multicolumn{2}{c||}{$\mu\mu$} & \multicolumn{3}{c||}{$\ell \ell$}\\ \cline{2-10}
 Process & SS & OS & SS& OS & SS & OS & SS &OS &Total\\
\hline
$t {\bar t} H$ (120 GeV)& &    &   &  &  &    &   &    &  \\
$H \rightarrow \tau \tau$& 49.6 &     103.2 &      26.3 &      46.6 &      27.8 &      51.8 &     103.7 &     201.6 &     305.3\\
$H \rightarrow  W W^*$&      81.6 &     173.0 &      40.4 &      86.7 &      41.2 &      89.5 &     163.2 &     349.2 &     512.4\\
\hline
\hline
$t {\bar t} H$ (125 GeV)& &    &   &  &  &    &   &    &  \\
$H \rightarrow \tau \tau$&44.6 &      82.7 &      21.2 &      42.2 &      20.9 &      43.1 &      86.7 &     167.9 &     254.6\\
$H \rightarrow  W W^*$&     116.3 &     245.0 &      57.6 &     121.2 &      59.4 &     123.7 &     233.4 &     489.9 &     723.3\\
\hline
\hline
$t {\bar t} H$ (130 GeV)& &    &   &  &  &    &   &    &  \\
$H \rightarrow \tau \tau$& 33.4 &      65.8 &      16.8 &      32.5 &      17.9 &      33.4 &      68.1 &     131.7 &     199.8\\
$H \rightarrow  W W^*$&     150.0 &     315.2 &      72.9 &     153.5 &      77.2 &     162.4 &     300.1 &     631.1 &     931.3\\
\hline
\hline
$t {\bar t}Z$ &     125.9 &     158.7 &      62.4 &     845.7 &      62.5 &     886.9 &     250.8 &    1891.2 &    2142.0\\
$WWZ$&  21.5 &     156.0 &      10.4 &     194.8 &      10.4 &     203.0 &      42.3 &     553.8 &     596.2\\
$ZZ$ & 228.6 &     474.9 &     116.6 &   34448.9 &     111.0 &   35783.0 &     456.2 &   70706.8 &   71163.0\\
$t {\bar t}$& 147.5 &  668973.8 &      98.3 &  334339.4 &      49.2 &  343632.1 &     295.0 & 1346945.3 & 1347240.3\\
$t {\bar t}j$&  3.5 &  502156.5 &       0.0 &  245773.5 &       3.5 &  255277.5 &       7.0 & 1003207.5 & 1003214.5\\
$t {\bar t}W$ &     471.6 &     920.8 &     223.9 &     450.2 &     244.9 &     458.4 &     940.4 &    1829.5 &    2769.9\\
%%%$t {\bar t} 2j$ & 0.0 &  258504.5 &       0.0 &  124918.4 &       0.0 &  137373.8 &       0.0 &  520796.7 &  520796.7\\
$Z2j$ &   0.0 &   36207.8 &       0.0 & 4900649.0 &       0.0 & 5019321.2 &       0.0 & 9956178.0 & 9956178.0\\
$Z3j$&   0.0 &    9668.3 &       0.0 & 1382073.3 &       0.0 & 1441322.7 &       0.0 & 2833064.2 & 2833064.2\\
$WWZj$ & 23.0 &     159.5 &      10.7 &     204.7 &      10.6 &     204.8 &      44.3 &     569.0 &     613.3\\
$ZZj$&   113.1 &     221.6 &      49.2 &   14930.5 &      49.2 &   15479.5 &     211.6 &   30631.6 &   30843.2\\
$ZZW$&   7.3 &       8.1 &       3.7 &      92.4 &       3.8 &      96.9 &      14.8 &     197.4 &     212.2\\
\hline
\hline
\end{tabular}
\vskip .3in
{\small Table 1: Number of Dilepton events  for 100 fb$^{-1}$ integrated luminosity.
The results for different flavor compositions with same-sign (SS) and opposite-sign (OS) 
electrons/muons are shown.
} }}
\end{center}
\vskip .3in

\begin{figure}[ht!]
\begin{center}
\raisebox{0.0cm}{\hbox{\includegraphics[angle=-90,scale=0.50]{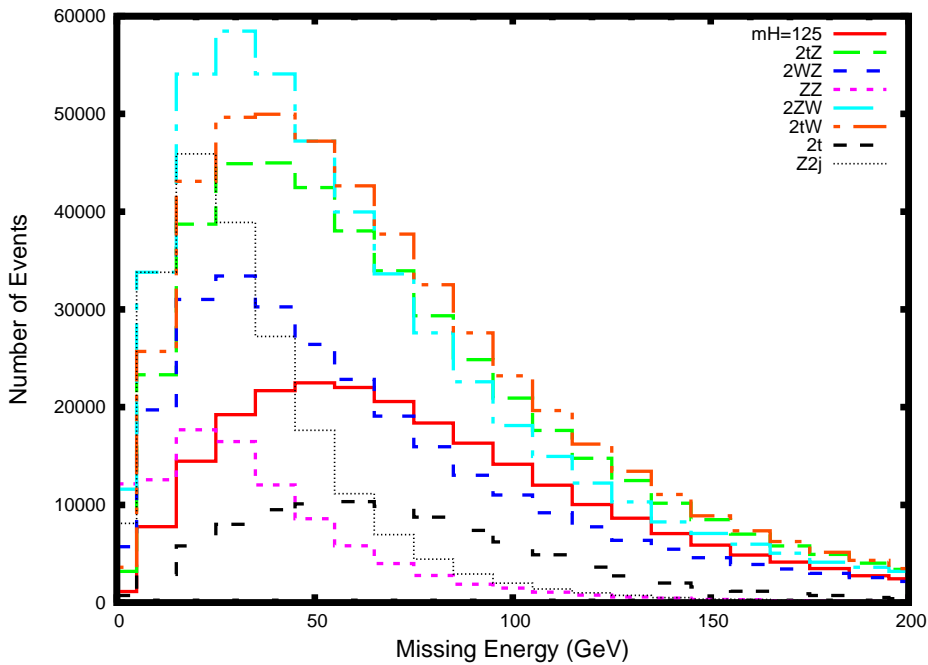}}}
\vskip .4in
\caption{\small Distribution of missing $p_T$  for the signal and the major SM backgrounds.} 
\label{met}
\end{center}
\end{figure}

	  In Table 2, we present the results for ``2 electrons/muons + a tau jet + a bottom-jet''. So
	  we wish to identify a bottom jet and a tau jet. We note that for the different masses 
	  of the Higgs boson, the number of signal events are almost identical. This is because
	  as  $M_H$ increases, the branching ratio $ H \to \tau \tau$ decreases, but
	  it increases for $H \to WW^{*}$. This together with different kinematics of the
	  electrons/muons from these two decay modes lead to nearly same events for different
	  $M_H$. For example, for the $M_H = 125$ GeV case,
          the contribution of the $WW^{*}$ decay mode is about $32 \%$, but for $M_H = 130$
	  GeV it is $60 \%$.
	  The signal events for this signature are the largest of all the considered signatures. 
	  This happens in part due to the appearance of only one tau jet. With 2 pairs
	  of W boson decaying into only three leptons, it gives rise to 
	  an additional combinatorial factor that increases the signal events. 
	  This signature has  very large background
	  from the $t {\bar t} W$ and $t {\bar t}$ processes. The significance
	  is not good for both the LTT and HTT cases. However, if we restrict
	  to the same-sign electrons/muons in the signature, the signature's
	  significance becomes more than 6, making it a pretty good signature.

	  In Table 3, we present the results for the signature 
	  ``2 electrons/muons + two tau jets''. The major backgrounds are
          $t {\bar t} Z, t {\bar t}, ZZ,$  and  $Z + 2 \;$ jets. Significance
	  for the 125 GeV Higgs boson is 4.0 for the HTT case. Because of
	  the reduction in the signal events, LTT case is not as useful. As
	  we see, restricting to the same-sign electrons/muons is again not
	  useful due to a paucity of events. We can also 
	  identify an additional bottom jet. This reduces the number of
	  signal events, but this also leads to a significant 
	  reduction in the Z boson backgrounds. As we see from Table 4,
	  this signature of ``2 electrons/muons + two tau jets + a bottom jet''
	  has a very good significance.

	  In the Table 5 we display the results for the signature
	  ``2 electrons/muons + a tau jet + two bottom jets''. Here
	  signal events are smaller as compared to that in Table 2.
	  This is due to the identification of an additional bottom jet.
	  As there, here the background due to the production of a top-quark pair 
	  is quite large. However, if we observe only the same-sign
	  electrons/muons, the significance may reach the observational value within
	  the run II of LHC.

\vskip 0.3in

\begin{center}
\begin{tabular}{||c|c|c|c|c|c|c|c|c|c|c||} \hline
 & \multicolumn{3}{c|}{Signal, $M_H$ (GeV)} & \multicolumn{4}{c|}{Backgrounds} & \multicolumn{3}{c||}{$S/\sqrt{B}$, $M_H$ (GeV)} \\ \cline{2-11}
  $\tau$ jets id & 120  & 125 & 130 & $t {\bar t} Z$ & $t {\bar t} $ & $t {\bar t} W$ &$Z2j$& 120  & 125 & 130 \\ \hline
LTT   & 333   & 333  & 330 & 336 & 8228 & 567 & 30 &3.4  & 3.4 & 3.4\\
HTT   & 555   &  552 & 549 & 561 & 32889 & 942 & 120 & 2.9 & 2.9 & 2.9  \\
SS/LTT  & 111  & 111 &  111 & 111 & 9 & 189 & 0 & 6.3 & 6.3 & 6.3 \\
SS/HTT  & 186 &  183 &  183 & 186 & 3 & 315 &0  & 8.3 &8.2 & 8.2\\ \hline
\end{tabular}
\vskip .3in
{\small
Table 2: Number of events for the signature ``2 electrons/muons + a tau jet + a bottom jet''
with the integrated luminosity of 300 fb$^{-1}$ with
the cuts and efficiencies specified in the text.}
\end{center}

\vskip 0.3in

\begin{center}
\begin{tabular}{||c|c|c|c|c|c|c|c|c|c|c|c|c||} \hline
 & \multicolumn{3}{c|}{Signal, $M_H$ (GeV)} & \multicolumn{6}{c|}{Backgrounds} & \multicolumn{3}{c||}{$S/\sqrt{B}$, $M_H$ (GeV)} \\ \cline{2-13}
  $\tau$ jets id & 120  & 125 & 130 & $t {\bar t} Z$ & $W W Z$ & $t {\bar t} W$ & $t {\bar t}$ & $Z2j$& $ZZ$ & 120  & 125 & 130 \\ \hline
LTT    & 42 & 41 & 37 & 36 & 6 & 3 & 9 & 9& 30& 4.4 & 4.3 & 3.8\\
HTT    & 117 & 114 & 104 &111 & 15  &9 & 147& 276&84 & 4.6 & 4.5 & 4.1\\
SS/LTT & 14  & 14  & 12 & 12 & 3   & 0& 0 & 0 & 0 & 3.6 & 3.6 & 3.1 \\
SS/HTT & 39 & 38 & 35 & 36 & 6 & 3& 0&0 & 0 & 5.8 & 5.7 & 5.2\\ \hline
\end{tabular}
\vskip .3in
{\small
Table 3: Number of events for the signature ``2 electrons/muons + 2 tau jets''
with the integrated luminosity of 300 fb$^{-1}$ with
the cuts and efficiencies specified in the text.}
\end{center}

\vskip 0.3in

\begin{center}
\begin{tabular}{||c|c|c|c|c|c|c|c|c|c||} \hline
 & \multicolumn{3}{c|}{Signal, $M_H$ (GeV)} & \multicolumn{3}{c|}{Backgrounds} & \multicolumn{3}{c||}{$S/\sqrt{B}$, $M_H$ (GeV)} \\ \cline{2-10}
  $\tau$ jets id & 120  & 125 & 130 & $t {\bar t} Z$ & $t {\bar t} W$ & $t {\bar t}j$ & 120  & 125 & 130 \\ \hline
LTT   & 34   & 33  & 30    &  30 & 3  & 6    & 5.4 & 5.3 & 4.8 \\
HTT   & 93   & 91  & 83   &  90  & 6 & 81  & 6.9  & 6.8 & 6.2\\
SS/LTT  & 11   & 11  & 10   &   10   & 0 & 0  & 3.5 & 3.5 &  3.2 \\
SS/HTT  & 31 & 30 & 28 & 30 & 3  & 0  &  5.4 & 5.2 & 4.9 \\ \hline
\end{tabular}
\vskip .3in
{\small
Table 4: Number of events for the signature ``2 electrons/muons + 2 tau jets + a bottom jet''
with the integrated luminosity of 300 fb$^{-1}$ with
the cuts and efficiencies specified in the text.}
\end{center}

\vskip 0.3in

\begin{center}
\begin{tabular}{||c|c|c|c|c|c|c|c|c|c|c||} \hline
 & \multicolumn{3}{c|}{Signal, $M_H$ (GeV)} & \multicolumn{3}{c|}{Backgrounds} & \multicolumn{3}{c||}{$S/\sqrt{B}$, $M_H$ (GeV)} \\ \cline{2-10}
  $\tau$ jets id & 120  & 125 & 130 & $t {\bar t} Z$ & $t {\bar t} j$& $t {\bar t} W$ &  120  & 125 & 130 \\ \hline
LTT   & 126   & 126  & 123 & 129 & 2286  & 213 & 2.4 & 2.4 & 2.4 \\
HTT   & 210  & 210  & 207 & 213 & 9141  & 357 & 2.1 & 2.1  & 2.1 \\
SS/LTT & 42 & 42  & 42  & 43 & 0  & 72 & 4.0 & 4.0 & 4.0 \\
SS/HTT  & 70 & 70  & 69 & 71 & 0  & 120 & 5.0 & 5.0 & 5.0 \\ \hline
\end{tabular}
\vskip .3in
{\small
Table 5: Number of events for the signature ``2 electrons/muons + a tau jet + two bottom jets''
with the integrated luminosity of 300 fb$^{-1}$ with
the cuts and efficiencies specified in the text.}
\end{center}

\vskip 0.3in

 Let us now comment on the possible uncertainties in the above results
 \cite{cmstth}.
 Theoretically, the main sources of uncertainties are choices of parton
 distribution functions, factorization and renormalization scales.
 In obtaining our results, we have used the NLO cross sections. These
 cross sections have the uncertainties of the order $10-15\%$.
 Furthermore, when these choices increase/decrease the signal
 cross section, they also correspondingly increase/decrease the background
 cross sections. Therefore, there is a further reduction in the
 uncertainties due to the cancellation when we compute the significance -- a ratio. Overall,
 one may expect only a few percent theoretical uncertainty in the
 significance of the signatures. Similarly, there will be cancellation
 of uncertainties due to experimental limitations. Therefore,
 our results about the significance are quite robust.

\section{Conclusion}

 In this letter, we have analyzed the signatures with two electrons/muons
 for the process $p p \to t {\bar t} H$. In particular, we have
 considered  the signatures ``2 electrons/muons + a tau jet + a bottom jet''
 ``2 electrons/muons + two tau jets'', ``2 electrons/muons + a tau jet + two bottom jets'',
 and ``2 electrons/muons + two tau jet + a bottom jet''. The major backgrounds are
 from the process $t {\bar t}$ (and jets) and the processes with Z bosons.
 The signatures with two tau jets have decent significance and may be observed
 in the run II of the LHC. The signatures with only one tau jet
 are overwhelmed by the backgrounds due a top-quark pair production. However,
 restricting to the same-sign electrons/muons events, these signatures
 may also be visible. So it appears that to observe the $t {\bar t} H$
 process using two electrons/muons, one may need to either tag two tau jets
 or tag one tau jet but observe same sign electrons/muons. More detailed 
 analysis of various other signatures will be presented elsewhere.

%{\bf {
%We use the same Gaussian energy smearing for PYCELL generated jets, with resolution
%%
%\bea \label{smearing}
%{\Delta E^{j} \over E^{j}} = {50\% \over \sqrt{E^{j}}}
%\oplus 3\% \quad ,
%\eea
%%
%where $\oplus$ means addition in quadrature. 
%
%For the leptons, the smearing is:
%%
%\bea \label{smearing}
%{\Delta E^{\ell} \over E^{\ell}} = {10\% \over \sqrt{E^{\ell}}}
%\oplus 0.7\% \quad .
%\eea
%

%The $p_T$, $\eta$ and isolation criterion imposed after smearing.
%}}

\bibliographystyle{plainnat}

\end{document}